\documentclass[twocolumn,english,pra,showpacs, twocolumn]{revtex4}
\usepackage[T1]{fontenc}
\usepackage[latin9]{inputenc}
\usepackage{babel}
\usepackage{amsmath}
\usepackage{amssymb}
\usepackage{graphicx}
\usepackage{esint}
\usepackage[unicode=true,pdfusetitle,
 bookmarks=true,bookmarksnumbered=false,bookmarksopen=false,
 breaklinks=false,pdfborder={0 0 1},backref=false,colorlinks=false]
 {hyperref}
\usepackage{breakurl}

\makeatletter
\@ifundefined{textcolor}{}
{%
 \definecolor{BLACK}{gray}{0}
 \definecolor{WHITE}{gray}{1}
 \definecolor{RED}{rgb}{1,0,0}
 \definecolor{GREEN}{rgb}{0,1,0}
 \definecolor{BLUE}{rgb}{0,0,1}
 \definecolor{CYAN}{cmyk}{1,0,0,0}
 \definecolor{MAGENTA}{cmyk}{0,1,0,0}
 \definecolor{YELLOW}{cmyk}{0,0,1,0}
}

\makeatother

\begin{document}

\title{Non-canonical statistics of finite quantum system}

\author{D. Z. Xu,$^{1,2,4}$ Sheng-Wen Li,$^{2,4}$ X. F. Liu,$^{3}$ and
C. P. Sun$^{2,4}$}

\email{cpsun@csrc.ac.cn}

\homepage{http://www.csrc.ac.cn/~suncp/}

\selectlanguage{english}%

\affiliation{$^{1}$State Key Laboratory of Theoretical Physics, Institute of
Theoretical Physics, Chinese Academy of Science, Beijing 100190, China\\
 $^{2}$Beijing Computational Science Research Center, Beijing 100084,
China\\
$^{3}$Department of Mathematics, Peking University, Beijing 100871,
China\\
$^{4}$Synergetic Innovation Center of Quantum Information and Quantum
Physics, University of Science and Technology of China, Hefei, Anhui
230026, China}
\begin{abstract}
The canonical statistics describes the statistical properties of an
open system by assuming its coupling with the heat bath infinitesimal
in comparison with the total energy in thermodynamic limit. In this
paper, we generally derive a non-canonical distribution for the open
system with a finite coupling to the heat bath, which deforms the
energy shell to effectively modify the conventional canonical way.
The obtained non-canonical distribution reflects the back action of
system on the bath, and thus depicts the statistical correlations
through energy fluctuations.
\end{abstract}


\maketitle

\section{introduction}

Statistical mechanics describes the average properties of a system
without referring its all microscopic states. In most situations,
the validity of the canonical statistical description is guaranteed
in the thermodynamic limit, which requires that, while the degrees
of freedom of the heat bath is infinite, the system-bath coupling
approaches to infinitesimal. However, if the system only interacts
with a small heat bath with finite degrees of freedom, the system-bath
interaction cannot be ignored. The properties of such finite system
recently intrigue a lot of attentions from the aspects of both experiments\cite{Weiss,Bustamante}
and theories \cite{Hill,Dong2007,Olshanii,JRau,HHasegawa1,HHasegawa2,ASarracino,WGWang,MEKellman}.

Although the canonical statistical distribution has been built on
a rigorous foundation \cite{Loinger,Tasaki,Popescu,Goldstein}, the
conventional canonical statistics still cannot well describe the thermodynamic
behavior of the finite system when the sufficiently large system-bath
interaction is taken into consideration \cite{Dong2007,Dong2010}.
To tackle this problem, we generally consider an effective system-bath
coupling by assuming the bath possesses a much more dense spectrum
than that of the system, then the system-bath interaction energy can
be treated as the deformation of the energy shell for the total system.
Therefore, the canonical distribution is modified to be a non-canonical
one with explicit expression. This modified distribution obviously
implies that corrections are necessary for the finite system thermodynamic
quantities in canonical statistics, such as average internal energy
and its fluctuation. 

The rest of the paper is organized as follows. In Sec. II we derive
an effective Hamiltonian of the total system by perturbation theory,
via this Hamiltonian the non-canonical statistical distribution without
referring to any specific model is presented. To further illustrate
the novel thermodynamic properties of the finite system by non-canonical
statistics, a model of coupled harmonic oscillators is introduced
in Sec. III, and the statistical quantities such as internal energy,
fluctuation and the mutual information between two subsystem are calculated.
We conclude in Sec. IV.

\section{finite system-bath coupling}

\begin{figure}
\includegraphics[width=7cm]{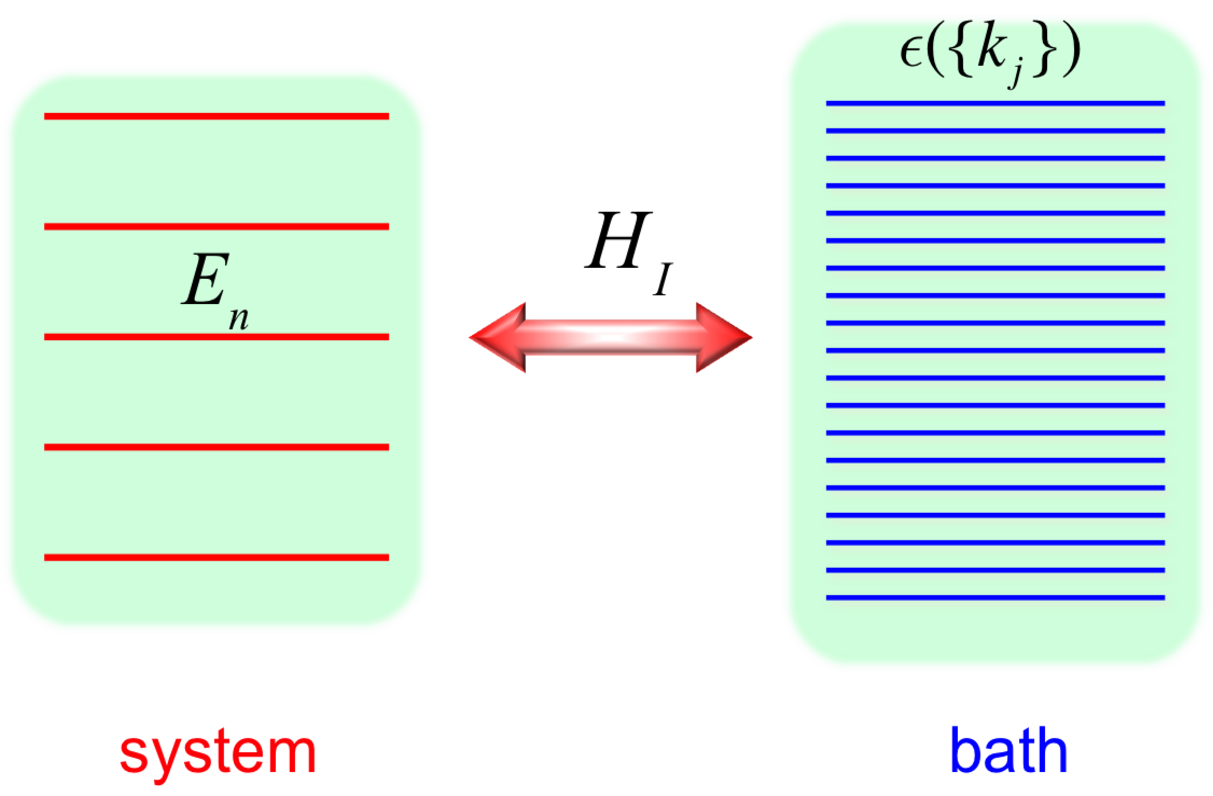}

\caption{\label{fig:EnergyLevels}The energy spectra of the system and the
heat bath. The spectral density of the heat bath should be much larger
then the system.}
\end{figure}
We generally consider a composite coupled system, which can be divided
into a system $S$ with Hamiltonian $H_{S}$ and a heat bath $B$
with Hamiltonian $H_{B}$. The coupling between the system and the
bath can be generally described by $H_{I}$. Then we have the total
Hamiltonian $H=H_{S}+H_{B}+H_{I}$. The system and the bath have the
following spectrum decompositions 
\begin{eqnarray}
H_{S} & = & \sum_{n}E_{n}\left|n\right\rangle \left\langle n\right|,\label{eq:Hs}\\
H_{B} & = & \sum_{j=1}^{N}\sum_{k_{j}}\epsilon_{k_{j}}\left|k_{j}\right\rangle \left\langle k_{j}\right|.\label{eq:Hb}
\end{eqnarray}
Here, $\left|n\right\rangle $ is the eigenstate of the system with
the corresponding eigen energy $E_{n}$. The heat bath is composed
of $N$ identical particles, the eigenstate of the $j$th particle
is $\left|k_{j}\right\rangle $ with the corresponding eigen energy
$\epsilon_{k_{j}}$. Usually, the energy spectrum of the system is
much sparser than the heat bath, i.e., 
\begin{equation}
|E_{n}-E_{m}|\gg|\epsilon(\{k_{j}\})-\epsilon(\{l_{j}\})|,\label{eq:condi}
\end{equation}
which holds for two arbitrary energy levels $n$ and $m$ of the system
(see Fig. \ref{fig:EnergyLevels}), and two arbitrary bath energy
states configurations $\{k_{j}\}$ and $\{l_{j}\}$. Here, we denote
$\epsilon(\{k_{j}\})=\sum_{\{k_{j}\}}\epsilon_{k_{j}}$.

The system-bath interaction $H_{I}$ is weak comparing to $H_{0}$,
which generally reads 
\begin{equation}
H_{I}=\sum_{j=1}^{N}\sum_{n,n^{\prime},k_{j},k_{j}^{\prime}}g_{nk_{j},n^{\prime}k_{j}^{\prime}}\left|nk_{j}\right\rangle \left\langle n^{\prime}k_{j}^{\prime}\right|\label{eq:V}
\end{equation}
with $\left|nk_{j}\right\rangle \equiv\left|n\right\rangle \otimes\left|k_{j}\right\rangle $.
Next we consider the role of the off-diagonal terms with respect to
the system indexes $n$ in $H_{I}$. The first order perturbation
effect of these off-diagonal terms with $n\neq n^{\prime}$ can be
ignored under the condition
\begin{equation}
\Big|\frac{g_{nk_{j},n^{\prime}k_{j}^{\prime}}}{E_{n^{\prime}}-E_{n}+\epsilon_{k_{j}^{\prime}}-\epsilon_{k_{j}}}\Big|\ll1,\label{eq:pertubation}
\end{equation}
However, for the terms with $n=n^{\prime}$, the above condition $\big|g_{nk_{j},nk_{j}^{\prime}}(\epsilon_{k_{j}^{\prime}}-\epsilon_{k_{j}})^{-1}\big|\ll1$
will be violated due to the properties of the energy spectra given
in Eq. (\ref{eq:condi}). Thus the diagonal terms can contribute to
the system behaviors and should be kept in the interaction Hamiltonian
\cite{Dong2010}, which yields 
\[
H_{I}\approx\sum_{j=1}^{N}\sum_{n,k_{j},k_{j}^{\prime}}g_{nk_{j},nk_{j}^{\prime}}\left|nk_{j}\right\rangle \left\langle nk_{j}^{\prime}\right|.
\]
Then, the total effective Hamiltonian has the diagonal form with respect
to the eigenstates of the system, i.e., 
\begin{equation}
H_{\mbox{eff}}=\sum_{n}[E_{n}+h(n)]\left|n\right\rangle \left\langle n\right|\label{eq:Heff}
\end{equation}
with 
\begin{equation}
h(n)=\sum_{j=1}^{N}\sum_{k_{j},k_{j}^{\prime}}\left(\epsilon_{k_{j}}\delta_{k_{j},k_{j}^{\prime}}+g_{nk_{j},nk_{j}^{\prime}}\right)\left|k_{j}\right\rangle \left\langle k_{j}^{\prime}\right|.\label{eq:hn}
\end{equation}
Here, $h(n)$ describes the heat bath Hamiltonian corresponding to
the system energy level $\left|n\right\rangle $. It can be further
diagonalized as $h(n)=\sum_{j=1}^{N}\sum_{\alpha_{j}}\varepsilon_{\alpha_{j}}(n)\left|\alpha_{j}\right\rangle \left\langle \alpha_{j}\right|$.
The new energy spectrum of the heat bath $\{\varepsilon_{\alpha_{j}}\}$
deforms comparing with the original energy spectrum $\{\epsilon_{k_{i}}\}$
due to the system-bath coupling. This system-bath coupling is usually
negligible when we study the thermalized state of the system in a
large heat bath. However, if the dimension of the heat bath is relatively
small, i.e. the finite system thermodynamic case, this coupling has
significant effect on modifying the canonical distribution of the
system. 

To derive the canonical distribution of the system in a large heat
bath, it is usually assumed an energy shell between $E-E_{n}$ and
$E-E_{n}+\delta$ in the phase space, where $E$ is the total energy
of the system and heat bath, $\delta$ is the thickness of the energy
shell which is a small quantity. This energy shell includes a set
of states
\begin{equation}
V_{0}\left(E-E_{n}\right):\{\left|n,\{k_{i}\}\right\rangle \vert E-E_{n}\leq\sum_{\{k_{j}\}}\epsilon_{k_{j}}\leq E-E_{n}+\delta\}.\label{eq:Ve0}
\end{equation}
Here, the system-bath interaction energy is neglected and $V_{0}$
gives a constrain on the configurations of heat bath states $\left|\{k_{i}\}\right\rangle $
when the system energy is fixed at $E_{n}$. Then, according to the
postulate that each microscopic state has an equal priori probability,
the probability of the system in the state $\left|n\right\rangle $
is proportional to the number of states in $V_{0}\left(E-E_{n}\right)$,
\begin{equation}
P_{0}(E,E_{n})=\frac{\Omega(E-E_{n})}{W(E)},\label{eq:PEn0}
\end{equation}
where $\Omega(E-E_{n})$ denotes the number of states satisfying the
constrain Eq.(\ref{eq:Ve0}), and $W(E)=\sum_{n}\Omega(E-E_{n})$
\cite{Huang1987}.

\begin{figure}
\includegraphics[width=8cm]{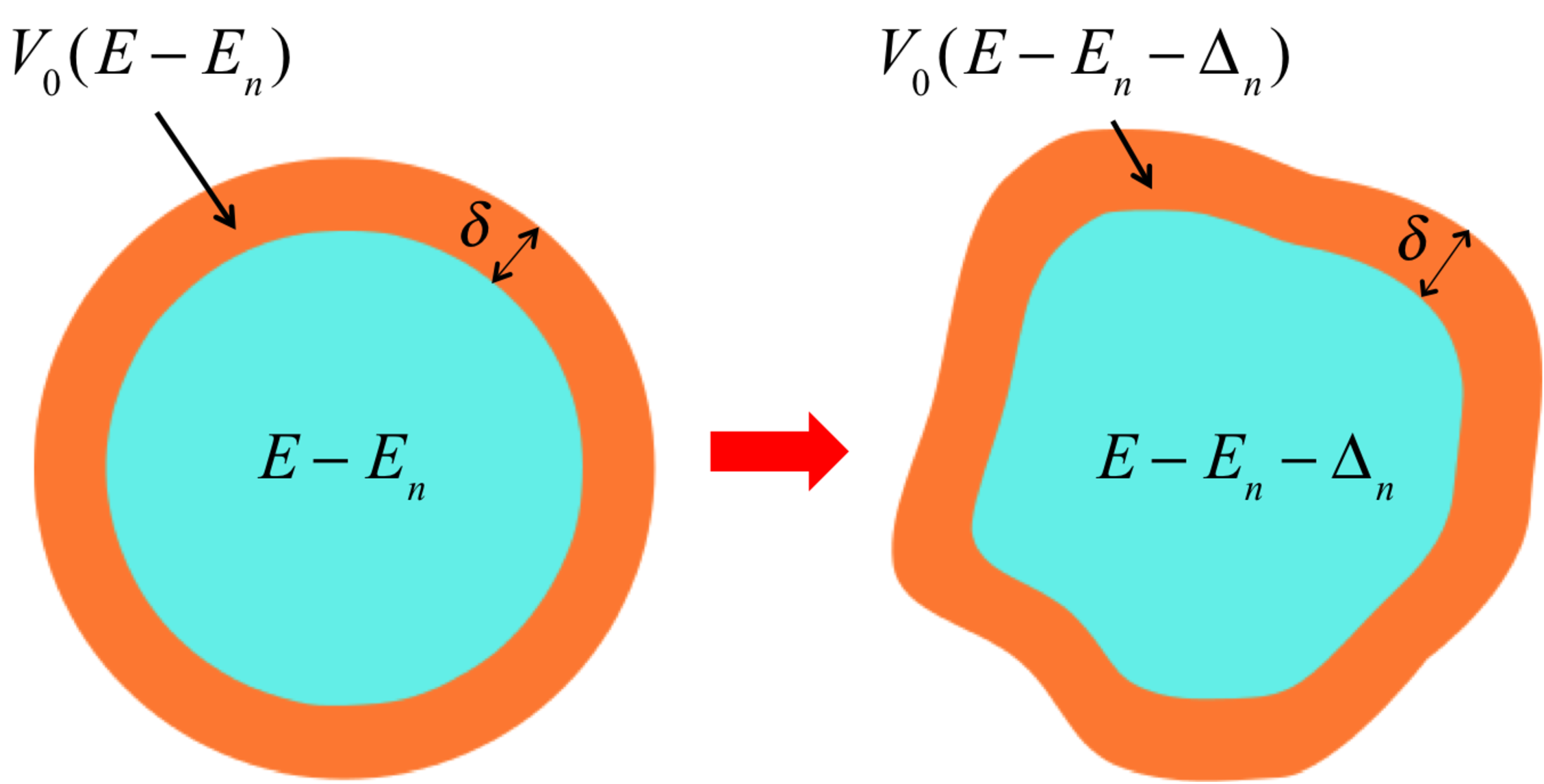}

\caption{\label{fig:EnergyShell}The energy shell without and with the consideration
of the system-bath coupling, which shifts the energy shell for $\Delta_{n}$. }
\end{figure}

In the situation of finite system statistics, it is crucial to consider
the system-bath interaction energy for the relatively small heat bath.
The effective Hamiltonian Eq. (\ref{eq:Heff}) is already diagonalized
with diagonal elements $E(n,\{\alpha_{i}\})=E_{n}+\sum_{\{\alpha_{i}\}}\varepsilon_{\alpha_{i}}(n)$.
It explicitly defines an energy shell as the subspace $V\left(E-E_{n}\right)$:
\begin{equation}
\{\left|n,\{\alpha_{i}(n)\}\right\rangle \vert E-E_{n}\leq\sum_{\{\alpha_{i}\}}\varepsilon_{\alpha_{i}}(n)\leq E-E_{n}+\delta\}.\label{eq:Ve}
\end{equation}
Nevertheless, it is more convenient to count the number of states
in $V\left(E-E_{n}\right)$ via the bath bases $\{\left|k_{i}\right\rangle \}$.
Because $V\left(E-E_{n}\right)$ is a geometrical deformed energy
shell comparing with $V_{0}\left(E-E_{n}\right)$, we re-express it
as $V_{0}\left(E-E_{n}-\Delta_{n}\right)$:
\begin{equation}
\{\left|n,\{\alpha_{i}(n)\}\right\rangle \vert E-E_{n}-\Delta_{n}\leq\sum_{\{k_{i}\}}\epsilon_{k_{i}}\leq E-E_{n}-\Delta_{n}+\delta\},\label{eq:VeD}
\end{equation}
where the geometrical deformation of the energy shell is characterized
by $\Delta_{n}=\sum_{\{\alpha_{i}\}}\varepsilon_{\alpha_{i}}(n)-\sum_{\{k_{i}\}}\epsilon_{k_{i}}$.
The deformation of the energy shell is schematically illustrated in
Fig. \ref{fig:EnergyShell}. By counting the number of states in $V_{0}\left(E-E_{n}-\Delta_{n}\right)$,
we can obtain the probability of the the system in the state $\left|n\right\rangle $
as 
\begin{equation}
P(E,E_{n})=\frac{\Omega(E-E_{n}-\Delta_{n})}{W(E)}.\label{eq:PEn}
\end{equation}
Different from $P_{0}(E,E_{n})$ in Eq. (\ref{eq:PEn0}), $P(E,E_{n})$
takes the system-bath coupling into account. 

To further obtain the statistic distribution of the system, we introduce
the entropy as $S(E)=k\ln\Omega(E)$, where $k$ is the Boltzmann
constant. Usually, $E_{n}+\Delta(n)$ is much smaller than the total
energy $E$, thus the entropy reads
\begin{eqnarray}
S(E-E_{n}-\Delta_{n}) & \approx & S(E)-\frac{\partial S(E^{\prime})}{\partial E^{\prime}}\bigg|_{E^{\prime}=E}(E_{n}+\Delta_{n})\nonumber \\
 &  & +\frac{\partial^{2}S(E^{\prime})}{\partial E^{\prime2}}\bigg|_{E^{\prime}=E}\Delta_{n}E_{n}\label{eq:S2}
\end{eqnarray}
The first term $S(E)$ is independent of $E_{n}$, thus it does not
determine the specific distribution form. The derivative terms of
$S(E)$ can be evaluated directly from its definition, as $\Omega(E)$
is proportional to the volume of the energy shell in the phase space.
In a $N$-dimensional space, the volume confined in an isoenergic
surface of energy $E$ can be considered as the volume of a $N$-dimensional
polyhedron with effective radius $E$, which is proportional to $E^{\zeta N}$,
here $\zeta$ is a dimensionless real number independent of $N$ and
is usually related to the degeneracy of the system states. Therefore,
the volume of the energy shell is given by 
\begin{equation}
\Omega(E)\varpropto(E+\delta)^{\zeta N}-E^{\zeta N},\label{eq:Omega}
\end{equation}
which leads to\emph{
\begin{equation}
\beta\left(E\right)=\frac{\partial S(E^{\prime})}{\partial E^{\prime}}\bigg|_{E^{\prime}=E}=\frac{k\left(\zeta N-1\right)}{E}.\label{eq:T}
\end{equation}
}With the thermodynamic relations, $\beta=(kT)^{-1}$ and $T$ is
the temperature of the finite system in equilibrium. Obviously, Eq.(\ref{eq:T})
recovers the equipartition theorem $E\approx\zeta NkT$, i.e., each
degree of freedom contributes $\zeta kT$ to the total energy. Therefore,
the non-canonical statistic distribution function reads
\begin{eqnarray}
P(E,E_{n}) & = & \frac{1}{\mathcal{Z}}e^{-\beta(E_{n}+\Delta_{n})+\xi\Delta_{n}E_{n}},\label{eq:PeN}
\end{eqnarray}
where $\mathcal{Z}=\sum\exp[-\beta(E_{n}+\Delta_{n})+\xi\Delta_{n}E_{n}]$
is the partition function with \emph{
\begin{equation}
\xi(E)=\frac{\partial^{2}S(E^{\prime})}{\partial E^{\prime2}}\bigg|_{E^{\prime}=E}=-\frac{k\left(\zeta N-1\right)}{E^{2}}.\label{eq:xi}
\end{equation}
}

If the interaction energy could be neglected comparing with the total
energy $E$, all the terms containing $\Delta_{n}$ in Eq. (\ref{eq:PeN})
can be dropped and thus naturally lead to the usual canonical statistics
distribution function
\begin{eqnarray}
P_{can}(E,E_{n}) & = & \frac{1}{\mathcal{Z}}\exp(-\beta E_{n}).\label{eq:Pcan}
\end{eqnarray}
Otherwise, the deformation of energy shell $\Delta_{n}$ will shift
the system energy level and thus modify the distribution function.
We can sort the system eigenvalues in an ascending order with $E_{n+1}>E_{n}$
for any $n$. Accordingly, we assume the energy shell deformation
$\Delta_{n}<0$ and $|\Delta_{n+1}|>|\Delta_{n}|$ (as illustrated
by the general model below). It follows from Eqs.(\ref{eq:PeN}, \ref{eq:Pcan})
that 
\begin{equation}
\frac{P(E,E_{n+1})}{P(E,E_{n})}>\frac{P_{can}(E,E_{n+1})}{P_{can}(E,E_{n})},\label{eq:comp}
\end{equation}
which means in the non-canonical statistics, the higher energy levels
play more important role than that in the canonical statistics.

\section{illustration with harmonic oscillators system}

Now we consider a coupled harmonic oscillators system as an example
to illustrate the statistical thermodynamic properties of a finite
system. The system $S$ is a single harmonic oscillator with eigenfrequency
$\omega$ and $a^{\dagger}(a)$ as its creation (annihilation) operator.
The heat bath is generally modeled as a collection of harmonic oscillators
with Hamiltonian $H_{B}=\sum_{j=1}^{N}\omega_{j}b_{j}^{\dagger}b_{j}$.
Here $b_{j}^{\dagger}(b_{j})$ is the creation (annihilation) operator
of the oscillator with frequency $\omega_{j}$. In the weak coupling
limit, we can assume the effective system-bath interaction as 

\begin{equation}
H_{I}=\sum_{j=1}^{N}\lambda_{j}a^{\dagger}a(b_{j}^{\dagger}+b_{j}).\label{eq:HIo}
\end{equation}

In this model, the eigenvalues of the total Hamiltonian are 
\begin{align}
E(n,\{m_{j}(n)\}) & =n\omega+\Delta_{n}+\sum_{\{m_{j}\}}m_{j}\omega_{j},\label{eq:En}
\end{align}
which corresponding to the eigenstates $\left|n,\{m_{j}(n)\}\right\rangle =\left|n\right\rangle \otimes\prod_{j=1}^{N}\left|m_{j}(n)\right\rangle $.
The eigenstate of the heat bath $\left|m_{j}(n)\right\rangle $ is
defined as a displaced Fock state $\left|m_{j}(n)\right\rangle =D(-\lambda_{j}n/\omega_{j})\left|m_{j}\right\rangle $,
with the displacement operator $D(\alpha_{j})=\exp(\alpha a_{j}^{\dagger}-\alpha^{\ast}a_{j})$.
Here, the deformation of the energy shell is described by an $n$-dependent
factor
\begin{equation}
\Delta_{n}=-\sum_{j=1}^{N}\frac{\lambda_{j}^{2}n^{2}}{\omega_{j}}\equiv-\kappa n^{2}.\label{eq:Dn}
\end{equation}
Due to this deformation, the non-canonical distribution function with
high order correction is given by

\begin{eqnarray}
P(E,n\omega) & = & \frac{1}{\mathcal{Z}}e^{-\beta(n\omega-n^{2}\kappa)-\xi\omega\kappa n^{3}}.\label{eq:Pno}
\end{eqnarray}
The square and cubic terms of $n$ in the exponent of $P(E,n\omega)$
greatly change the statistical distribution from the canonical distribution
$P_{n}=\mathcal{Z}^{-1}\exp(-\beta n\omega)$, especially for large
$n$. However, Eq.(\ref{eq:Pno}) does not apply to very large $N$
for the following reason: the total energy $E$ of the system and
the heat bath is conserved, and the heat bath energy should always
be nonnegative $\sum_{\{m_{j}\}}m_{j}\omega_{j}>0$. Thus, the energy
shell of Eq.(\ref{eq:VeD}) constrains the system energy level by
$E\geq E_{n}$, which implies $n<(\omega-\sqrt{\omega^{2}-4\kappa E})/(2\kappa)$.
Therefore, for such coupled finite system, the maximum energy level
for the system is 
\begin{equation}
n_{max}=\left\lfloor \frac{\omega-\sqrt{\omega^{2}-4\kappa E}}{2\kappa}\right\rfloor ,\label{eq:nmax}
\end{equation}
here $\left\lfloor x\right\rfloor $ represents the maximum integer
below $x$.

\begin{figure}
\includegraphics[width=7cm]{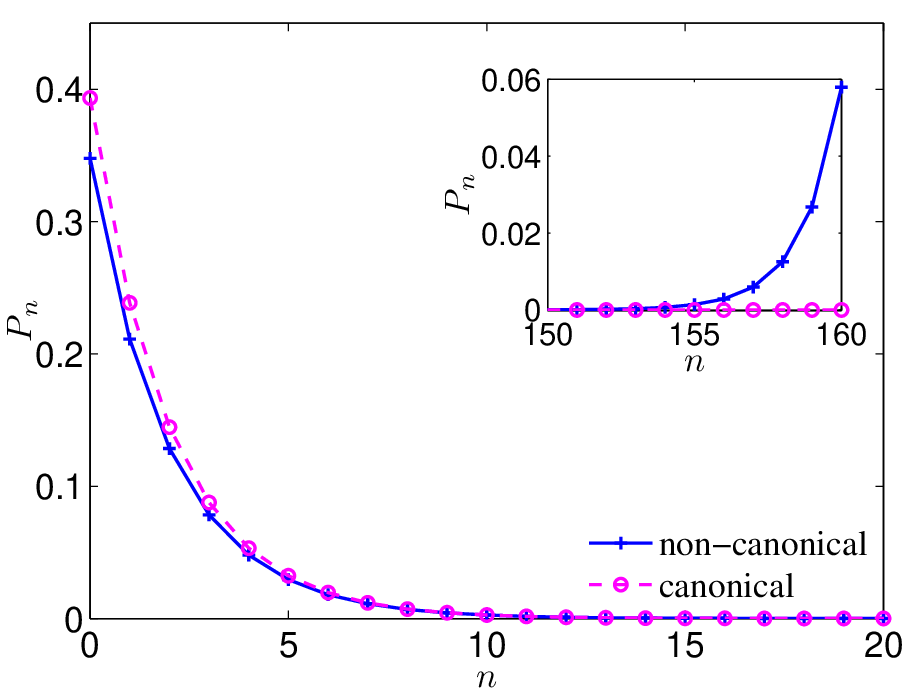}

\caption{\label{fig:Pn}The non-canonical (blue solid line) and canonical (pink
dashed line) distribution functions for coupled oscillators system.
We choose $\omega=1$, $E=100$, $\beta=0.5$ and $\kappa=0.00235$.
The highest energy level is $n_{max}=160$. The non-canonical and
canonical distributions are similar for low energy levels while of
great difference for high energy levels.}
\end{figure}

Usually $\kappa$ is much smaller than $\omega$, thus the non-canonical
statistics distribution $P(E,n\omega)$ for low energy levels is not
very different from the canonical one, as shown in Fig. \ref{fig:Pn}.
However, as $|\Delta_{n}|$ grows with $n^{2}$, the high energy levels
share much more ratio than that in the canonical distribution, which
can be seen from the inset of Fig. \ref{fig:Pn}. We choose the system
eigen frequency as unit $\omega=1$, the total energy $E=100$ and
$\kappa=0.00235$. According to Eq.(\ref{eq:nmax}), the highest energy
level is $n_{max}=160$ in this situation. The canonical distribution
is plotted by set $\kappa=0$ in $P(E,n\omega)$. A numerical research
also gave the similar non-canonical distribution for coupled spin
systems \cite{WXZhang2010}.

\begin{figure}
\includegraphics[width=7cm]{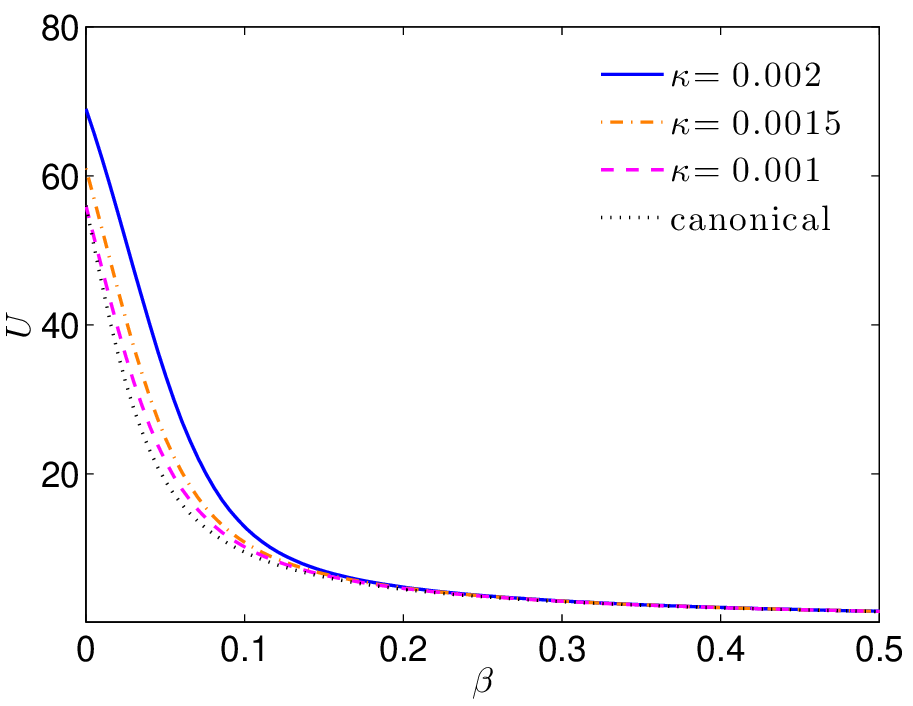}

\caption{\label{fig:U}The internal energy $U$ of the system with respect
to $\beta$ for non-canonical and canonical statistics. We plot the
cases of $\kappa=0.002$, $n_{max}=138$ (blue solid line), $\kappa=0.0015$,
$n_{max}=122$ (orange dotted-dash line), $\kappa=0.001$, $n_{max}=112$
(pink dashed line) and the canonical case for $\kappa=0$, $n_{max}=112$. }
\end{figure}

Because the high energy states have relatively larger populations,
the internal energy of the system 
\begin{equation}
U=\sum_{n=0}^{n_{max}}n\omega P(E,n\omega)\label{eq:U}
\end{equation}
under the non-canonical statistics is larger than that under the canonical
one. This fact is illustrated in Fig. \ref{fig:U}, where the internal
energy $U$ is plotted with respect to $\beta$. The distinction between
the non-canonical and canonical statistics for $U$ evidently appears
when the inverse temperature $\beta$ decreases and the interaction
energy strength $\kappa$ grows. As $\beta$ approaches to zero, the
high temperature limit of the internal energy $U$ arrives at $n_{max}\omega/2$,
which is finite as the total energy is up-bounded by $E$ for small
heat bath. This is very different from the case in the thermodynamic
limit: the average energy of a harmonic oscillator which contacts
with an infinite heat bath will diverge when $\beta$ decreases to
zero.

\begin{figure}
\includegraphics[width=7cm]{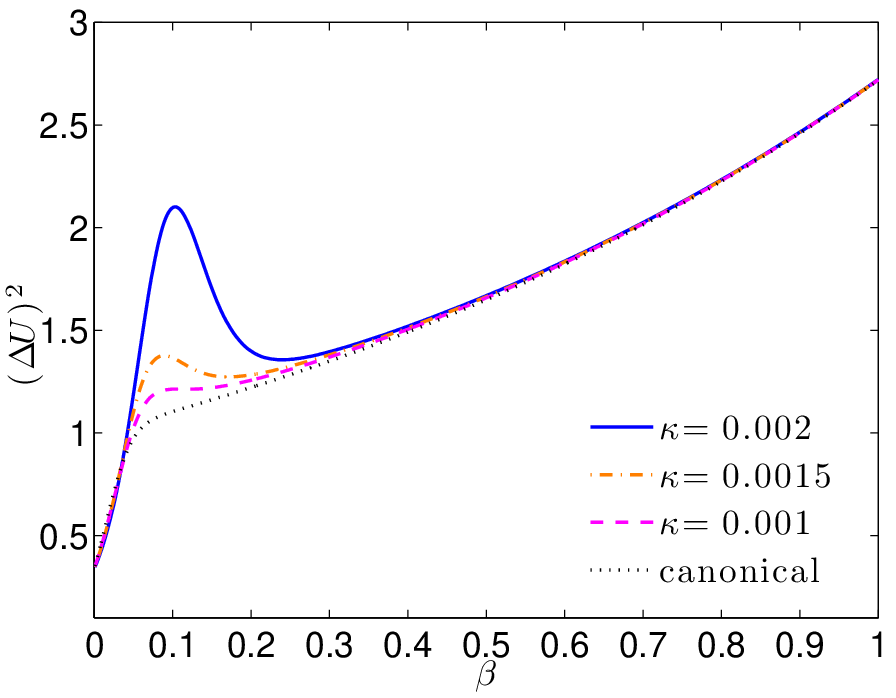}

\caption{\label{fig:F}The internal energy relative fluctuation $(\Delta U)^{2}$
of the system with respect to $\beta$ for non-canonical and canonical
statistics. The parameters are chosen the same as Fig. \ref{fig:U}.}
\end{figure}

Another feature reflecting the non-monotony of the non-canonical distribution
is the relative fluctuation of the system internal energy 
\begin{equation}
\left(\Delta U\right)^{2}=\frac{1}{U^{2}}\sum_{n=0}^{n_{max}}\left(n\omega-U\right)^{2}P(E,n\omega).\label{eq:F}
\end{equation}
As shown in Fig. \ref{fig:F}, at both low and high temperature limits,
the non-canonical and canonical statistics for finite system present
similar fluctuation behavior. To characterize these two limits, we
consider the system as a harmonic oscillator with truncated energy
levels (the highest energy level is labeled by $n_{max}$) under canonical
statistics ($\kappa=0$), whose energy fluctuation is denoted as $(\Delta U_{C})^{2}$.
It is analytically calculated that in the low temperature limits,
the fluctuation $(\Delta U_{C})^{2}\approx\exp(\beta\omega)$ is exactly
the same as the result when $n_{max}\rightarrow\infty$. Because the
populations of the high energy levels decrease significantly in the
low temperature, only the several low energy levels determine the
thermodynamic behavior. In the high temperature limit, the energy
fluctuation behaves as 
\[
(\Delta U_{C})^{2}\approx\frac{1}{3}+\frac{2}{3n_{max}}+\frac{n_{max}}{9}\beta,
\]
which is a linear function of $\beta$. We remark here that the high
temperature limit and thermodynamic limit cannot commute with each
other, as
\[
\lim_{n_{max}\rightarrow\infty}\lim_{\beta\rightarrow0}(\Delta U_{C})^{2}=\frac{1}{3},
\]
while 
\[
\lim_{\beta\rightarrow0}\lim_{n_{max}\rightarrow\infty}(\Delta U_{C})^{2}=1.
\]

However, in the intermediate range of $\beta$, a local maximum in
energy fluctuation distinguishes the non-canonical distribution from
the canonical one, especially for strong system-bath interaction $\kappa$.
This maximum can be qualitatively understood as follow: we can rewrite
the non-canonical distribution as $P(E,n\omega)=\mathcal{Z}^{-1}\exp(-\beta\eta_{n}n\omega)$,
where $\eta_{n}=1-\kappa n/\omega-\kappa n^{2}/E$ is a positive factor
for $n\leq n_{max}$. As $\eta_{n}\leq1$, the reverse temperature
$\beta$ can be considered as effectively reduced by $\eta_{n}$,
thus the linear region for small $\beta$ is enlarged in the non-canonical
statistics. Based on the above observations, we know that the non-canonical
statistics exhibits obviously novel effects when the interaction energy
strength $\kappa$ is large and the temperature is high. 

\begin{figure}
\includegraphics[width=7cm]{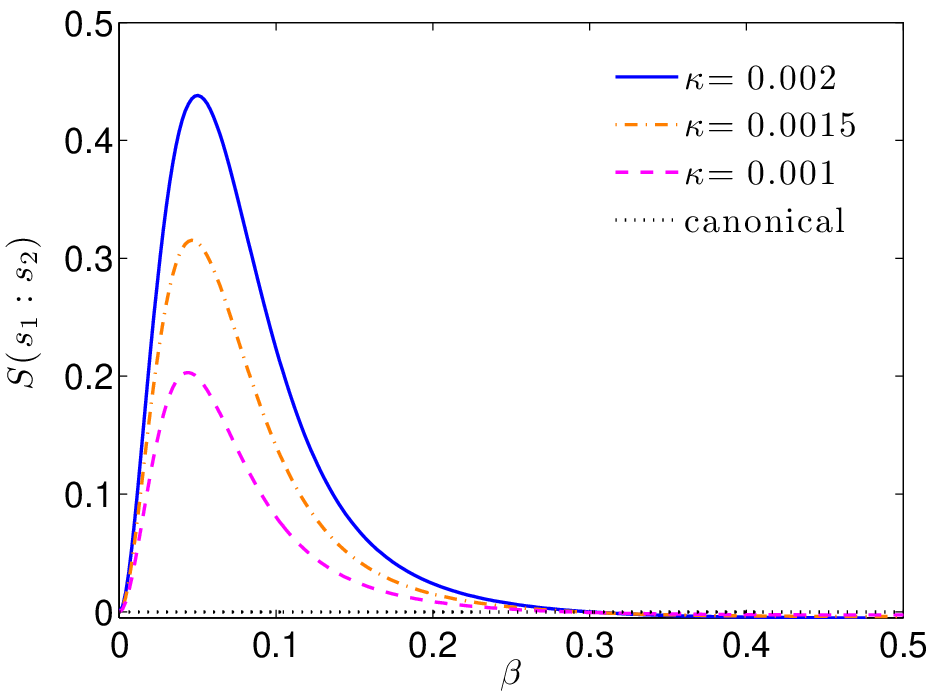}

\caption{\label{fig:M-S}The mutual entropy of two identical harmonic oscillators
(HOs) by non-canonical and canonical statistics. The eigen frequency
of two identical HOs is $\omega=1$, and $\kappa=0.002$, $n_{max}=138$
(blue solid line), $\kappa=0.0015$, $n_{max}=122$ (orange dotted-dash
line), $\kappa=0.001$, $n_{max}=112$ (pink dashed line) and the
canonical case for $\kappa=0$, $n_{max}=112$. }
\end{figure}

Besides the high distribution tail for a single system, the non-canonical
statistics provides other new characters when the system is composite
of two independent subsystem $l_{1}$ and $l_{2}$. Even if these
two subsystems do not directly interact with each other, the deformation
of the energy shell can effectively result in correlation between
them. Here we still use harmonic oscillator (HO) systems for illustration.
The system consists of two single mode HOs with Hamiltonian $H_{S}=\sum_{k=1,2}\omega_{k}a_{k}^{\dagger}a_{k}$.
The system interacts with a common small heat bath, which can be modeled
by the Hamiltonian $H_{B}=\sum_{j=1}^{N}\omega_{j}b_{j}^{\dagger}b_{j}$.
The interaction term reads
\[
H_{I}=\sum_{k=1,2}\sum_{j=1}^{N}\lambda_{kj}a_{k}^{\dagger}a_{k}(b_{j}^{\dagger}+b_{j}).
\]
Following the same discussion about the energy shell deformation for
a single system, we can straightforwardly obtain the joint distribution
of the composite system as
\begin{eqnarray}
P(E,E_{n}^{(1)},E_{m}^{(2)}) & = & \frac{1}{\mathcal{Z}_{t}}e^{-\beta(E_{nm}+\Delta_{nm})+\xi\Delta_{nm}E_{nm}},\label{eq:Pm}
\end{eqnarray}
where $E_{nm}=n\omega_{1}+m\omega_{2}$ and 
\begin{eqnarray*}
\Delta_{nm} & = & -\sum_{j=1}^{N}\frac{(\lambda_{1j}n+\lambda_{2j}m)^{2}}{\omega_{j}}\\
\mathcal{Z}_{t} & = & \sum_{n,m}\,^{\prime}e^{-\beta(E_{nm}+\Delta_{nm})+\xi\Delta_{nm}E_{nm}}.
\end{eqnarray*}
Here $\sum_{n,m}^{\prime}$ means the summation of the system energy
levels should satisfy the constrain $0\leq E_{nm}+\Delta_{nm}\leq E$.
It can be seen from Eq.(\ref{eq:Pm}) that the statistics of two subsystems
are not independent with each other due to the cross-term in $\Delta_{nm}$.
This statistical correlation can be described by mutual information
defined as
\[
S(l_{1}:l_{2})=S(l_{1})+S(l_{2})-S(l_{1}+l_{2}),
\]
where the entropy $S(l_{k})=-\sum_{n=0}^{n_{max}}P(E,E_{n}^{(k)})\ln P(E,E_{n}^{(k)})$
and $S(l_{1}+l_{2})=-\sum_{n,m}^{\prime}P(E,E_{n}^{(1)},E_{m}^{(2)})\ln P(E,E_{n}^{(1)},E_{m}^{(2)})$.
For simplicity, we assume the two HOs are identical with $\omega_{1}=\omega_{2}\equiv\omega$,
$\lambda_{1j}=\lambda_{2j}\equiv\lambda_{j}$ and $\kappa$ is defined
the same as Eq.(\ref{eq:Dn}). As shown in Fig. \ref{fig:M-S}, there
appears a non-zero mutual entropy if we use non-canonical statistics
to describe the composite system in a common small heat bath. In contrary,
if the interaction energy is too small to be considered comparing
with the total energy, we can use the canonical distribution to calculate
the mutual entropy which naturally gives $S(l_{1}:l_{2})=0$, i.e.,
the two subsystems are not correlated with each other.

\section{conclusion}

We study the \textit{\emph{statistical thermodynamics of an open system
whose interaction with the heat bath cannot be neglect. The interaction
modifies the system energy shell and leads to the non-canonical statistical
distribution for such system. It is shown that non-canonical distribution
has a big ``tail'' for higher energy levels, which is the most significant
difference from the canonical distribution. This non-canonical feature
results in higher internal energy and energy fluctuation of the system.
And different parts of the composite system are naturally correlated
with each other which is described by mutual entropy. We would like
to mention that the non-canonical form of distribution may be related
to the explanation of black hole information paradox \cite{MSZhan}. }}

\emph{This work is supported by National Natural Science under Grants
No.11121403, No. 10935010, No. 11074261, No.11222430, No.11074305
and National 973 program under Grants No. 2012CB922104.}


\begin{thebibliography}{10}
\bibitem{Weiss} T. Kinoshita, T. Wenger, and D. S. Weiss, Nature
(London) \textbf{440}, 900 (2006).

\bibitem{Bustamante} J. Liphardt, S. Dumont, S. B. Smith, I. Tinoco
Jr., C. Bustamante, Science \textbf{296}, 1832 (2002).

\bibitem{Hill} T. L. Hill, J. Chem. Phys. \textbf{36}, 3182 (1962).

\bibitem{Dong2007} H. Dong, S. Yang, X. F. Liu, and C. P. Sun, Phys.
Rev. A \textbf{76}, 044104 (2007).

\bibitem{Olshanii} M. Rigol, V. Dunjko, and M. Olshanii, Nature (London)
\textbf{452}, 854 (2008).

\bibitem{JRau} J. Rau, Phys. Rev. A \textbf{84}, 012101 (2011).

\bibitem{HHasegawa1} H. Hasegawa, Phys. Rev. E \textbf{83}, 021104
(2011).

\bibitem{HHasegawa2} H. Hasegawa, J. Math. Phys. \textbf{52}, 123301
(2011). 

\bibitem{ASarracino} M. Falcioni, D. Villamaina, A. Vulpiani, A.
Puglisi, and A. Sarracino, Am. J. Phys. \textbf{79}, 777 (2011).

\bibitem{WGWang} W. G. Wang, Phys. Rev. E \textbf{86}, 011115 (2012).

\bibitem{MEKellman} G. L. Barnes and M. E. Kellman, J. Chem. Phys.
\textbf{139}, 214108 (2013). 

\bibitem{Loinger} P. Bocchieri and A. Loinger, Phys. Rev. \textbf{114},
948 (1959).

\bibitem{Tasaki} H. Tasaki, Phys. Rev. Lett. \textbf{80}, 1373 (1998).

\bibitem{Popescu} S. Popescu, A. J. Short, and A. Winter, Nat. Phys.
\textbf{2}, 754 (2006). 

\bibitem{Goldstein} S. Goldstein, J. L. Lebowitz, R. Tumulka, and
N. Zanghì, Phys. Rev. Lett. \textbf{96}, 050403 (2006).

\bibitem{Dong2010} H. Dong, X.F. Liu, and C. P. Sun, Chin. Sci. Bull.
\textbf{55}, 3256 (2010).

\bibitem{Huang1987} K. Huang, Statistical Mechanics, John Wiley \&
Sons (1987).

\bibitem{WXZhang2010} Wenxian Zhang, C. P. Sun, and Franco Nori,
Phys. Rev. E \textbf{82}, 041127 (2010).

\bibitem{MSZhan} B. Zhang, Q. Y, Cai, L. You, and M. S. Zhan, Phys.
Lett. B \textbf{675}, 98 (2009); B. Zhang, Q. Y, Cai, M. S. Zhan,
and L. You, Ann. Phys. \textbf{326}, 350 (2011).\end{thebibliography}
\end{document}